\documentclass[%
superscriptaddress,
preprint,
nofootinbib,
 amsmath,amssymb,
 aps,
pra,
]{revtex4-2}

\usepackage{xcolor}
\usepackage{subfigure}
\usepackage{graphicx}
\usepackage{dcolumn}
\usepackage{bm}

\usepackage{epstopdf}
\usepackage[utf8]{inputenc}
\usepackage[T1]{fontenc}

\begin{document}

\title{Temperature-Driven Evolution of Magnetization Dynamics in Cobalt-Doped \textbf{$Fe_{5}GeTe_{2}$}}

\author{Sourabh Singh}
\thanks{These authors contributed equally to this work. \textbf{Contact}: s.singh-2@utwente.nl (S.S.), soumik.aon2@gmail.com (S.A.)}
\affiliation{NanoElectronics Group, MESA+ Institute, and BRAINS Center for Brain-Inspired Computing, University of Twente, PO Box 217, Enschede 7500 AE, The Netherlands.}

\author{Soumik Aon}
\thanks{These authors contributed equally to this work. \textbf{Contact}: s.singh-2@utwente.nl (S.S.), soumik.aon2@gmail.com (S.A.)}
\affiliation{Department of Physics, National University of Singapore, 2 Science Drive 3, Singapore 117551, Republic of Singapore}

\author{Andrew F. May}
\affiliation{Materials Science and Technology Division, Oak Ridge National Laboratory, Oak Ridge, Tennessee 37832, USA}

\author{Joost Ridderbos}
\affiliation{NanoElectronics Group, MESA+ Institute, University of Twente, PO Box 217, Enschede 7500 AE, The Netherlands.}

\author{Ramanathan Mahendiran}
\affiliation{Department of Physics, National University of Singapore, 2 Science Drive 3, Singapore 117551, Republic of Singapore}

\author{Wilfred G. van der Wiel}
\affiliation{NanoElectronics Group, MESA+ Institute, and BRAINS Center for Brain-Inspired Computing, University of Twente, PO Box 217, Enschede 7500 AE, The Netherlands.}
\affiliation{Institute of Physics, University of Münster, 48149 Münster, Germany}

\begin{abstract}
Understanding magnetization dynamics in metallic van der Waals ferromagnets is essential to uncover their magnetic interactions and evaluate their potential for high-frequency spintronics applications. In this study, we investigate the temperature-dependent magnetization dynamics of a single crystal of cobalt-doped $Fe_{5}GeTe_{2}$ (CFGT) using broadband ferromagnetic resonance (FMR) spectroscopy. At high temperatures, the FMR response is dominated by a single broad resonance that shifts toward lower magnetic fields upon cooling. Fits to the frequency dependence of the resonance field and linewidth yield an effective $g$ factor of approximately 2.45 and an effective Gilbert damping parameter of approximately 0.15 near room temperature. The effective magnetic anisotropy evolves strongly with temperature, with an increasingly pronounced easy-plane contribution upon cooling toward 200 K. Below approximately 200 K, an additional resonance emerges and evolves into a distinct two-mode response that persists to the lowest measured temperature of 25 K. The two branches exhibit different temperature dependences of their resonance fields and linewidths, and the emergence of the additional mode is not accompanied by a comparably pronounced anomaly in the static magnetization. These results reveal a dynamical crossover in CFGT that is not readily apparent from bulk magnetometry and highlight the sensitivity of FMR to the magnetic behavior of chemically substituted van der Waals ferromagnets.
\end{abstract} 
\maketitle
   \section{\label{sec:level1}Introduction}
Layered van der Waals (vdW) magnetic materials have emerged as a promising class of systems for investigating the interplay between dimensionality, exchange interactions, and magnetic anisotropy  \cite{hou2025magnetic}\cite{wu2025spin}. Their weak interlayer bonding allows individual layers to be isolated and the materials to be thinned toward the two-dimensional limit. Unlike conventional magnetic materials, vdW materials can be integrated into heterostructures without the constraints imposed by lattice matching, making them attractive both for investigating low-dimensional magnetism and for developing nanoscale spin-based devices \cite{yu2026realizing}. In addition to fundamental questions regarding magnetic ordering and spin fluctuations, understanding magnetization dynamics in these materials is essential for assessing their suitability for high-frequency and low-power spintronics applications \cite{kajale2024two}. In particular, magnetic anisotropy, effective magnetization, and magnetic damping govern the response to microwave excitation, influencing the resonance characteristics, relaxation processes, and ultimately the performance of spin-based devices  \cite{tang2023spin}.

Among metallic vdW ferromagnets, $Fe_{5}GeTe_{2}$ (F5GT) has emerged as a particularly interesting system owing to its high Curie temperature, complex magnetic anisotropy, and rich magnetic phase behavior \cite{may2019ferromagnetism}\cite{deng2022layer}. Unlike simpler layered ferromagnets, Fe$_5$GeTe$_2$ contains multiple inequivalent Fe sites, giving rise to complex exchange interactions and a strong coupling between structural and magnetic degrees of freedom \cite{alahmed2021magnetism}. These characteristics have motivated extensive investigations of its magnetic, electronic, and structural properties \cite{schmitt2022skyrmionic}. In addition, its metallic character makes F5GT attractive for spintronics applications, where efficient electrical manipulation of magnetization and an understanding of high-frequency spin dynamics are particularly important \cite{liu2022layer}\cite{bindu2025quantum}.

Chemical substitution offers an effective means of tuning the magnetic properties of F5GT by modifying its local magnetic environments and exchange interactions \cite{hu2024tuning}. Previous studies have demonstrated that transition-metal substitution can substantially alter the magnetic ground state, anisotropy, and thermodynamic properties of F5GT based compounds \cite{birch2024influence} \cite{ma2025tuning}. The crystal structure of F5GT consists of Fe-Ge slabs sandwiched between Te layers and contains three crystallographically inequivalent Fe sites, Fe1, Fe2, and Fe3. Among these, Fe1 occupies a partially occupied split site, which has been associated with the complex magnetic behavior of the parent compound. Cobalt (Co) preferentially occupies Fe1 sites and has been reported to stabilize a rhombohedral structure with ABC stacking, thereby modifying local magnetic environments and magnetic anisotropy \cite{may2020tuning}. 

These structural modifications are accompanied by substantial changes in the magnetic properties. For a Co concentration of $x \approx 0.28$, the magnetic easy axis changes from out-of-plane in the parent compound to in-plane,  while the Curie temperature is enhanced and the ferromagnetic order is maintained above room temperature \cite{may2020tuning}. Furthermore, the low-temperature magnetic anomaly observed near 120 K in pristine F5GT is suppressed at this Co concentration, a behavior that has been associated with preferential Co occupation of the partially occupied Fe1 sites and the resulting modification of Fe1-related magnetic ordering. The absence of a corresponding anomaly in static magnetization suggests that the magnetic behavior differs substantially from that of the parent compound \cite{sampaio2025dzyaloshinskii}. Unlike substitution-induced suppression of magnetic order reported in some related vdW ferromagnets, Co-doped in $Fe_{5}GeTe_{2}$ (CFGT) provides a route to stabilizing an easy-plane ferromagnetic state with an enhanced ordering temperature \cite{sampaio2025dzyaloshinskii}. The combination of above-room-temperature ferromagnetism, metallic conductivity, and in-plane magnetic anisotropy makes this system particularly attractive for investigating microwave-frequency magnetization dynamics. Nevertheless, despite extensive studies of its static magnetic properties, the temperature-dependent spin dynamics and magnetic relaxation processes of (CFGT) remain comparatively unexplored. 

Ferromagnetic resonance (FMR) spectroscopy is a powerful technique for probing dynamical magnetic properties. Through measurements of the resonance condition and linewidth over a broad frequency range, FMR enables determination of the spectroscopic g factor, effective magnetization, magnetic anisotropy fields, and Gilbert damping parameter. The temperature evolution of these quantities provides direct insight into spin relaxation mechanisms, magnetic fluctuations, and possible changes in the magnetic state. Consequently, broadband FMR is particularly well suited for investigating chemically tuned magnetic materials in which competing interactions may generate nontrivial temperature-dependent behavior. This technique also allows for a comprehensive investigation of the spin structures associated with low-energy spin dynamics and variations in magnetic anisotropy \cite{bera2024anisotropic}. 

In this work, we investigate the magnetization dynamics of a single crystal of CFGT with an approximate Co concentration of 28$\%$ using broadband FMR spectroscopy between 350 and 25 K. From frequency-dependent resonance measurements, we extract the gyromagnetic ratio, effective magnetization, and Gilbert damping parameter as a function of temperature. In addition to the overall evolution of these dynamical parameters, we observe a distinct low-temperature anomaly emerging below approximately 200 K, indicating a modification of the magnetic response that is not readily apparent from the static magnetization behavior alone. Our results provide new insight into the spin dynamics of CFGT and highlight the sensitivity of broadband FMR to emergent magnetic phenomena in layered metallic ferromagnets.

\subsection{\label{sec:level2}Experimental Details}
\begin{figure*}
\centering
\includegraphics[width=1.0\textwidth]{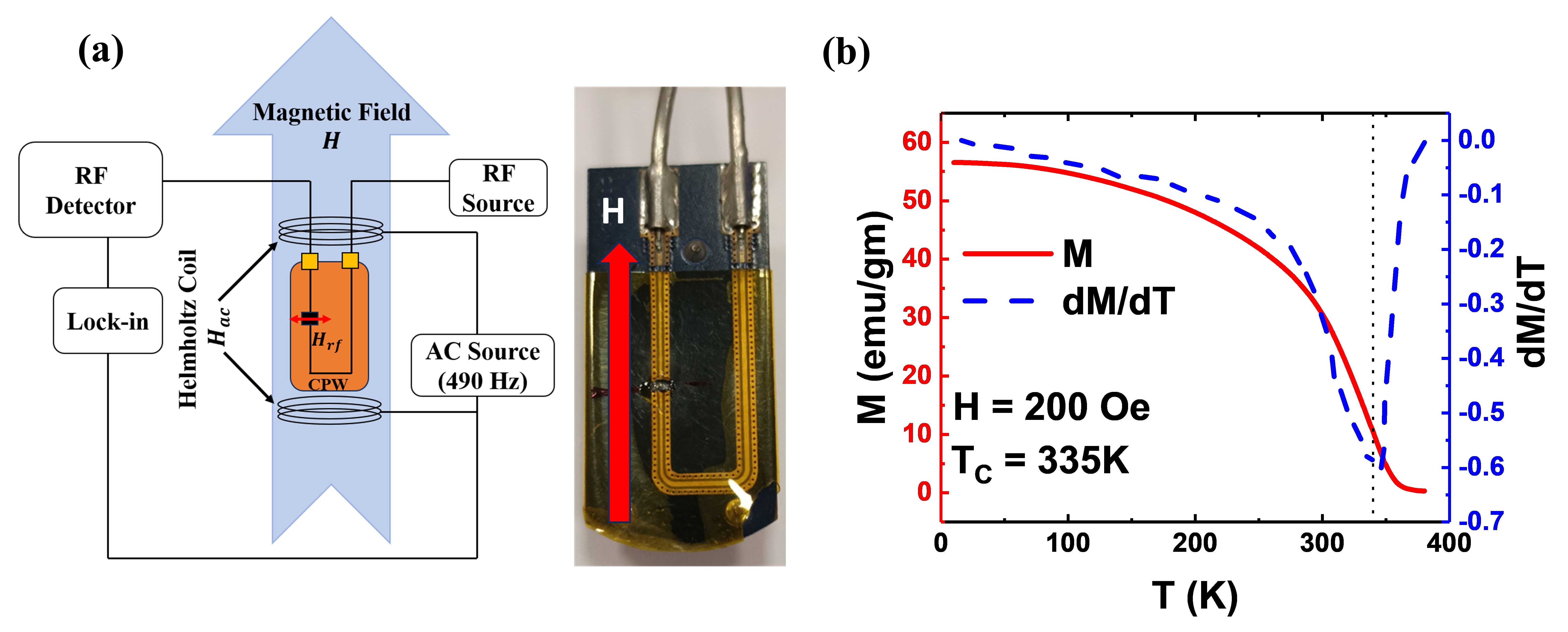}
\caption{(a) Schematic of the experimental design for the CPW-based cryo-FMR (left), two port in-plane CPW, where the sample is placed on the signal line (right), (b) static magnetization (M) data along with its temperature derivative (dM/dT) at a dc magnetic field H = 200 Oe. The minimum of dM/dT gives the Curie temperature $T_{C}$ = 335 K. Isothermal $M(H)$ curves are shown in Fig. S1 of the Supplementary Information.}
\label{Figure 1}
\end{figure*}

 Single crystals of CFGT were grown by a chemical vapor transport method using iodine as the transport agent, following the procedure described by May \textit{et al.}~\cite{may2020tuning}. High-purity elemental Fe, Co, Ge, and Te with nominal compositions
Fe$_{5-y}$Co$_y$GeTe$_2$ were sealed with iodine in evacuated silica ampoules. The ampoules were heated to 750~$^{\circ}$C at a rate of 120~$^{\circ}$C\,h$^{-1}$ and kept at this temperature for 1-2 weeks, after which they were quenched directly in ice water. Residual iodine on the crystal surfaces was removed with ethanol and/or isopropanol, followed by a final rinse in acetone. The crystals investigated in the present work were taken from the same growth batch as those characterized in Ref.~\cite{may2020tuning}, where the Co concentration was determined by energy-dispersive X-ray spectroscopy (EDS); the crystals used for the FMR measurements correspond to a Co substitution level of approximately 28\%. Both static magnetization (M) and ferromagnetic resonance (FMR) were measured in a PPMS system (Evercool, Quantum Design) equipped with 9T superconducting magnet. Vibrating sample magnetometer (VSM) was used to measure the static magnetization, with the sample placed such that the external magnetic field (H) is directed along the sample plane as shown in Fig. \ref{Figure 1}(a). The static magnetization data were taken during cooling with a rate of 4 K/min. For FMR, a NanOsc cryo-FMR broadband spectrometer was used. The sample was placed on top of a two-port coplanar waveguide (CPW) within the signal line with a static magnetic field applied along the sample plane. The FMR measurements were performed by sweeping the external magnetic field from the high-field saturation state toward zero, with a field step of 3 Oe. The sample was excited by microwave field with a working frequency range of 2-17 GHz, while the microwave power was kept constant throughout the measurement. A small ac field modulation of 490 Hz was applied through a pair of Helmholtz coils, and the transmitted power was detected and recorded by a lock-in amplifier, so that the measured signal is the field derivative of the transmitted power (dP/dH). The RF magnetic field ($H_{rf}$) was oriented perpendicular to the external dc field.  

\subsection{\label{sec:level3}Results and Discussion}

Fig. 1(a) illustrates the cryo-FMR measurement setup, showing the schematic of the measurement configuration (left) and the CPW (right) employed in this work. The single-crystalline CFGT sample is positioned on top of the CPW in the presence of an in-plane dc magnetic field, while the microwave magnetic field ($H_{rf}$) excites the magnetization dynamics. The FMR response was measured over a frequency range of 2-17 GHz using field modulation and lock-in detection, which gives the field-derivative signal dP/dH. Details of the experimental configuration are discussed in the Experimental section. Fig. 1(b) shows the M measured at an in-plane H = 200 Oe while cooling, together with its temperature derivative dM/dT. M decreases monotonically with increasing temperature and falls sharply near the magnetic transition. The minimum of dM/dT evaluates the Curie temperature ($T_{C}$) = 335 K. This is higher than the 310 K reported for undoped F5GT \cite{may2019ferromagnetism}, consistent with the enhancement of $T_{C}$ by Co substitution \cite{may2020tuning,tian2020tunable}. The high $T_C$ allows the FMR response to be studied over a wide temperature range.

The magnetization dynamics of ferromagnetic (FM) is described by the Landau–Lifshitz-
Gilbert (LLG) equation \cite{landau1935theory,gilbert2004}.
\begin{equation}\label{eq:1}
\frac{d\vec{M}}{dt} = -\gamma(\vec{M}\times\vec{H}_{eff}) + \frac{\alpha}{M_{s}}(\vec{M}\times\frac{d\vec{M}}{dt}) 
\end{equation}
The first term in Eq. \ref{eq:1} is the conservative precessional term, describing the precession of $\vec{M}$ about the effective magnetic field $\vec{H}_{eff}$. The second term is the phenomenological Gilbert damping term,
in which the dimensionless damping constant $\alpha$ quantifies the rate at which the precessing magnetization relaxes toward $\vec{H}_{eff}$. Here, $\gamma$ = $g\mu_{B}/\hbar$ is the gyromagnetic ratio, $g$ is the Lande $g$-factor, $\mu_{B}$ is the Bohr magneton, and $M_{s}$ is the saturation magnetization. The effective field $\vec{H}_{eff}$ consists of  $\vec{H}$ together with the demagnetizing, magneto-crystalline anisotropy, and exchange fields.\\
\begin{figure*}
\centering
\includegraphics[width=1.0\textwidth]{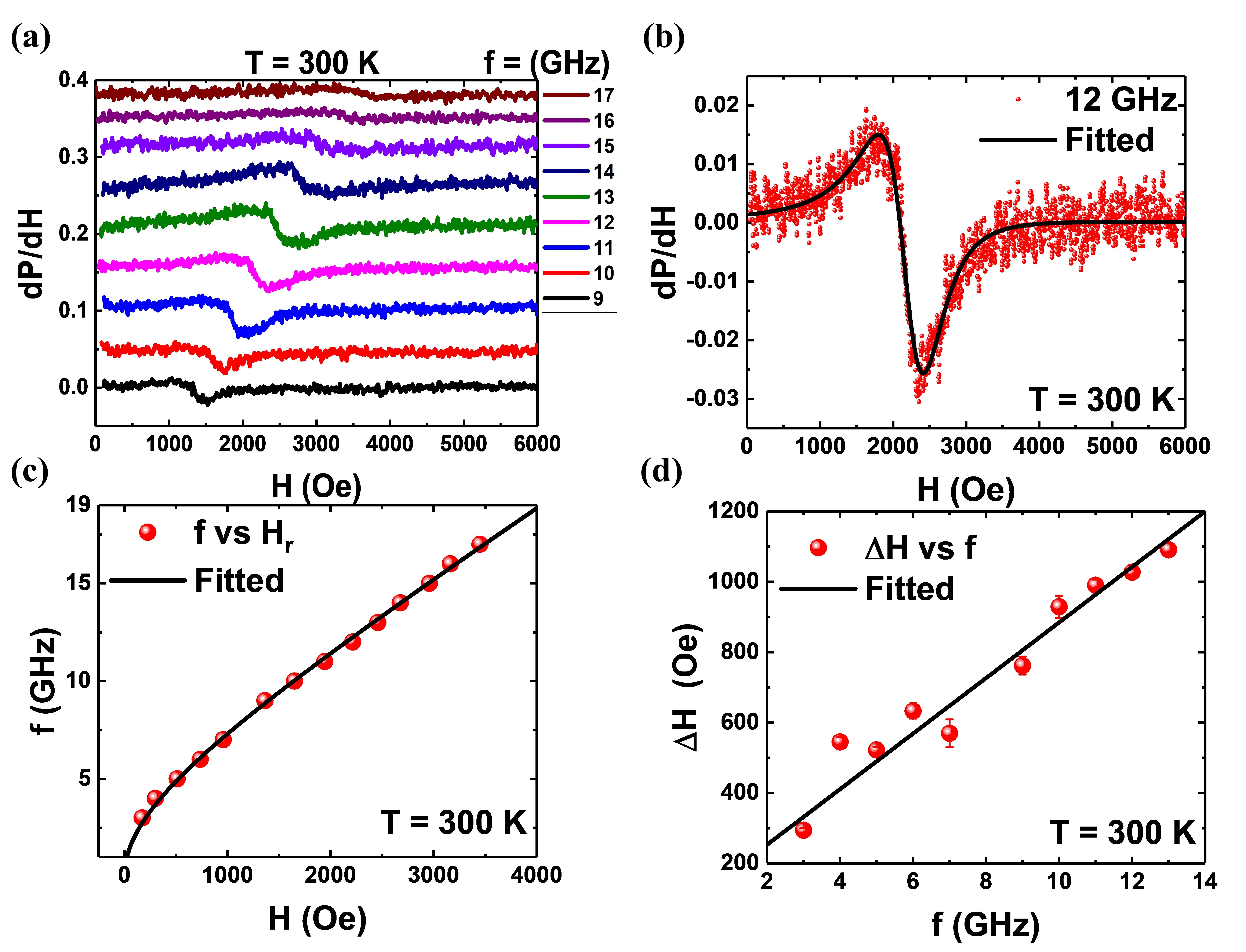}
\caption{(a) Field derivative of FMR response at room temperature over a frequency range 9-17 GHz, (b) The FMR spectrum at a frequency of 12 GHz is fitted using the symmetric and antisymmetric Lorentzian components. Red circles depict the experimental data points, whereas the black line represents the fitted plot, (c) Frequency versus $H_{r}$ is fitted with the Kittel relation, from which effective $g$-factor and $4\pi M_{eff}$ are evaluated, (d) $\Delta H$ increases linearly with frequency, and from the slope Gilbert damping $\alpha$ is evaluated.}
\label{Figure 2}
\end{figure*}
In our measurements, the FMR response is recorded by sweeping an in-plane magnetic field at fixed microwave frequencies. Figure 2(a) shows the room-temperature FMR spectra measured in the frequency range of 9-17 GHz. A single resonance is observed throughout this range, with the resonance field ($H_{r}$) shifting systematically to higher values with increasing frequency. Since field modulation and lock-in detection are used, the measured signal is the field derivative of the microwave response $dP/dH$. Each spectrum is therefore fitted with field derivative of a line-shape containing symmetric (absorptive) and antisymmetric (dispersive) Lorentzian functions \cite{harder2011}, together with a linear background: 
\begin{multline}\label{eq:2}
\frac{dP}{dH} = {P_{sym}}\frac{4\Delta H(H-H_{r})}{{[4(H-H_{r})^{2}+(\Delta H)^{2}]}^{2}} + {P_{asym}}\frac{[(\Delta H)^{2}-{4(H-H_{r})}^{2}]}{{[4(H-H_{r})^{2}+(\Delta H)^{2}]}^{2}} \\+ slope \times H + C
\end{multline}
Where, $P_{sym}$ and $P_{asym}$ are the symmetric and antisymmetric amplitude of the Lorentzian components. $H_{r}$ is the resonance field and $\Delta H$ is the linewidth. Linear term accounts for the slowly varying background and C is the constant. With increasing frequency, $H_{r}$ shifts to higher fields and $\Delta H$ increases, as expected for Gilbert-type damping \cite{kalarickal2006}. A representative fit is shown in Fig. \ref{Figure 2}(b). The frequency (f) dependence of $H_{r}$ is well described by the in-plane Kittel relation \cite{kittel1948,farle1998}:
\begin{equation}\label{eq:3}
f= \frac{g_{eff}\mu_{B}}{h}\sqrt{(H_{r}(H_{r}+4\pi M_{eff})}   
\end{equation}
Where, $M_{eff}$ is the effective magnetization, which combines the demagnetizing field with any perpendicular magnetic anisotropy field. The effective Lande $g$-factor ($g_{eff}$) and $4 \pi M_{eff}$ are evaluated from the Kittel relation [Eq. \ref{eq:3}], where, $g_{eff}$ = 2.45 $\pm$ 0.03 and $4\pi M_{eff}$ = 3515 $\pm$ 335 Oe. The linewidth increases linearly with $f$, as described by Eq. \ref{eq:4} \cite{kalarickal2006}:
\begin{equation}\label{eq:4}
\Delta H= \frac{2h\alpha} {g\mu_{B}}f + \Delta H_{0}
\end{equation}
A linear fit of $\Delta H$ versus $f$ [Eq. \ref{eq:4}] gives a Gilbert damping parameter $\alpha = 0.15$ and inhomogeneous linewidth broadening $\Delta H_{0}$ = 63 Oe [Plot given in Supplementary Information (SI) Fig. S4]. $\Delta H_{0}$ may arise from the local magnetic environment, structural disorder, or compositional inhomogeneity \cite{kalarickal2006}. Such inhomogeneity is expected in CFGT, whose structure contains three inequivalent Fe sites (Fe1, Fe2 and Fe3) \cite{may2019ferromagnetism} that are preferentially substituted by Co \cite{may2020tuning}. Both $g_{eff}$ and $\alpha$ are large compared with those of conventional 3$d$ ferromagnets. The value $g_{eff} = 2.45$ lies well above both the free-electron value (2.0023) and the typical value of $g_{eff} \approx 2.1$ for 3$d$ metals \cite{shaw2013}, suggesting a sizable orbital contribution to the magnetic moment mediated by spin orbit coupling \cite{kittel1949}. Because $g_{eff}$ and $M_{eff}$ are correlated when the Kittel relation is fitted over a limited frequency range \cite{shaw2013}, this $g_{eff}$ should be regarded an effective value. Similarly, $\alpha = 0.15$ may include extrinsic contributions in addition to intrinsic Gilbert relaxation. These observations highlight the complex magnetization dynamics of CFGT, while further measurements would be required to identify the dominant microscopic relaxation mechanisms.
\begin{figure*}
\centering
\includegraphics[width=1.0\textwidth]{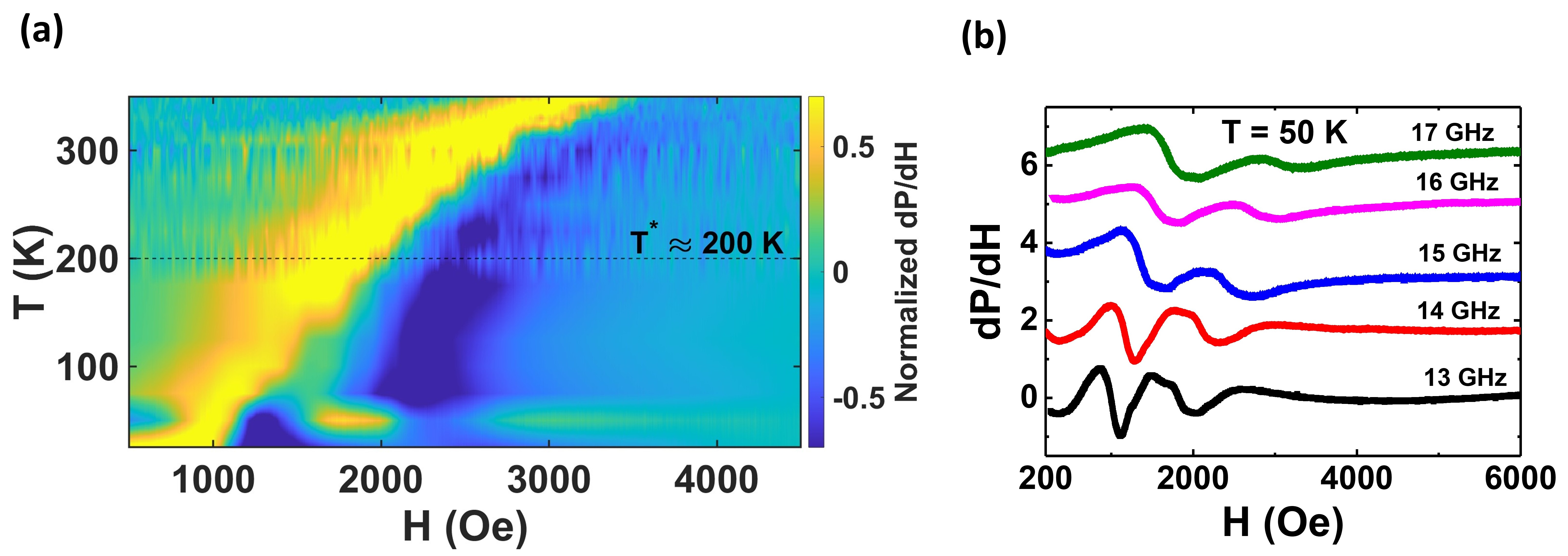}
\caption{Temperature and frequency evolution of the FMR response. (a) Temperature-field map of the normalized field-derivative FMR signal measured at 14 GHz. The dashed line marks $T^{*}\approx 200$ K, below which additional structure develops in the FMR response. The color map is interpolated between the measured temperatures for visualization. Each FMR spectrum was independently normalized to its maximum absolute signal amplitude at the corresponding temperature. The resulting normalized spectra were interpolated between the measured temperatures solely for visualization. Consequently, the color scale represents the relative spectral response at each temperature rather than the absolute temperature dependence of the FMR signal intensity.(b) Representative field-swept FMR spectra measured at 50 K for frequencies from 13 to 17 GHz. The spectra are vertically offset for clarity and reveal multiple frequency-dependent resonance features in the low-temperature regime.}
\label{fig:FMR_temperature}
\end{figure*}

The FMR response evolves considerably upon cooling, as summarized by the temperature–field map in Fig. \ref{fig:FMR_temperature}(a). At high temperatures, the spectra are dominated by a single broad resonance whose position shifts continuously toward lower magnetic fields with decreasing temperature. A marked change in the FMR response occurs below a characteristic temperature $T^{*}$ $\sim $ 200 K, where an additional resonance structure develops and becomes progressively more pronounced upon further cooling. The persistence and systematic evolution of these features indicate that the low-temperature response can no longer be described by a single uniform resonance mode. To investigate their frequency dependence, \ref{fig:FMR_temperature}(b) shows representative field-swept $dP/dH$ spectra measured at 50 K between 13 and 17 GHz. Multiple resonance features are clearly observed and shift systematically with microwave frequency, demonstrating that they represent dispersive magnetic excitations rather than a frequency-independent spectral feature. The low-temperature multimode response was reproducibly observed over multiple thermal cycles.
Interestingly, the appearance of this additional dynamical structure around $T^{*}$ is not accompanied by a comparably pronounced anomaly in the static magnetization. This suggests that the change in the FMR response should not simply be identified with a conventional bulk magnetic transition. Instead, the measurements point toward a more subtle reorganization of the magnetization dynamics below approximately 200 K. In particular, the composition $x\approx 0.28$  does not exhibit the pronounced static low-temperature magnetic transition reported for undoped F5GT, whereas the microwave response reveals a clear crossover into a multimode regime. FMR therefore provides sensitivity to changes in the dynamical magnetic environment that are not readily apparent from bulk magnetometry. The absence of a comparable feature in static magnetization therefore argues against simply identifying the additional FMR mode with the low-temperature transition of CFGT. The additional FMR mode observed below $T^*\approx200$ K should be distinguished from the low-temperature magnetic anomaly reported near 120 K in pristine F5GT, which is suppressed upon Co substitution at $x\approx0.28$ \cite{may2020tuning}. Unlike the anomaly associated with the parent compound, the emergence of the additional resonance in CFGT is not accompanied by a comparably pronounced feature in the static magnetization. Therefore, FMR can reveal changes in the dynamical magnetic environment that might not be visible through bulk magnetometry alone. The important observation is that the crossover is dynamical: it appears clearly in the microwave response even though the $x \approx 0.28$ composition does not show the pronounced low-temperature static magnetic transition characteristic of undoped F5GT.

Multiple resonance modes can originate from different mechanisms, including  spatially inhomogeneous magnetic environments, domain-related excitations, and nonuniform spin-wave modes  \cite{mcmichael2003localized} \cite{dkabrowski2021canted}. In CFGT, the appearance of an additional resonance is particularly intriguing given the complex magnetic structure of $Fe_{5}GeTe_{2}$, which contains three crystallographically inequivalent Fe sites (Fe1, Fe2, and Fe3), and the reported site preference of Co substitution \cite{may2020tuning}. These inequivalent sites possess distinct local coordination and exchange environments, potentially leading to different temperature-dependent magnetic responses. A redistribution of the dynamical response among magnetically inequivalent environments, or the development of additional collective excitation modes associated with their coupling, could therefore contribute to the observed multimode FMR behavior. Importantly, such a response would not necessarily require a separate bulk magnetic phase transition, consistent with the absence of a pronounced anomaly in the static magnetization near 200 K. Nevertheless, the present measurements do not uniquely identify the microscopic origin of the additional mode, and contributions from magnetic domains, spatial inhomogeneity, or nonuniform spin-wave excitations cannot be excluded.

\begin{figure*}
\centering
\includegraphics[width=1.0\textwidth]{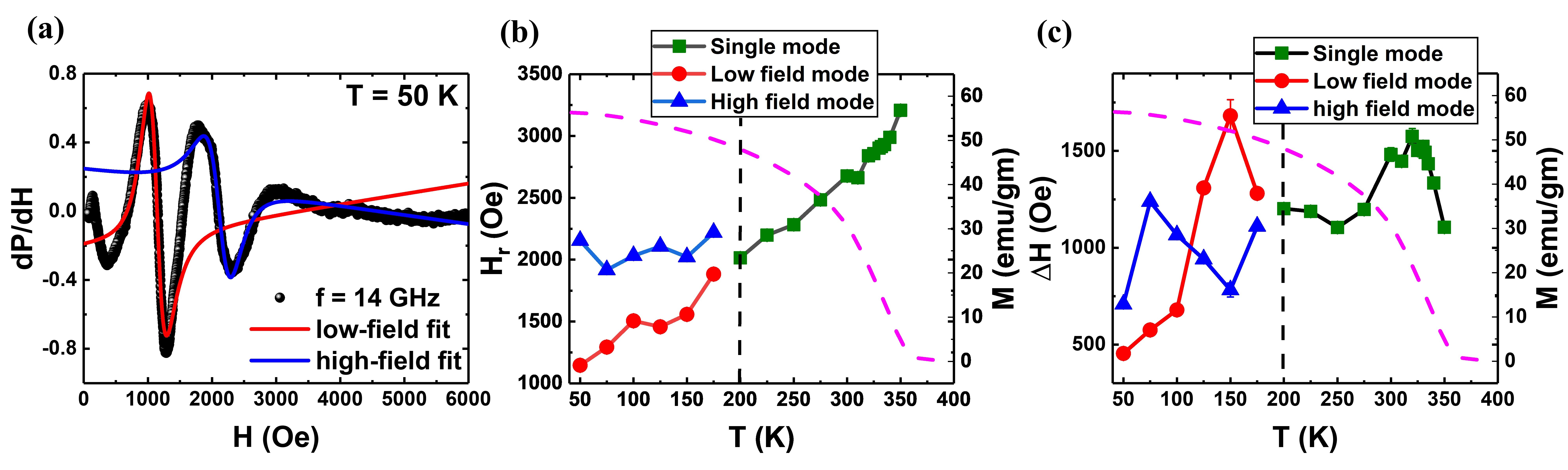}
\caption{Quantitative analysis of the temperature-dependent FMR response. (a) FMR spectrum at 50 K and 14 GHz with individual single line-shape fits to the low-field (red) and high-field (blue) features. (b) Temperature dependence of the resonance field $(H_{r})$, Above $T^* \approx 200$ K, the spectra are described by a single dominant resonance, whereas below $T^{*}$ distinct low- and high-field resonance branches are resolved. The dashed magenta curve shows the temperature-dependent magnetization $(M(T)$ [measured at 200 Oe] for comparison. (c) Temperature dependence of the corresponding FMR linewidth $\Delta H$ for the single-mode and low- and high-field resonance branches. The vertical dashed lines in (b) and (c) indicate $T^* \approx 200$ K, below which the multimode FMR response becomes evident. Points at 340 and 350~K ($T > T_C$) are effective parameters of a field-polarized state (see text).}
\label{Figure 4}
\end{figure*}

Further evidence for the multimode character of the low-temperature response is provided by the frequency-dependent spectra in Fig. 3(b). At 50 K, multiple resonance features are resolved in the measured frequency range of 13 to 17 GHz and shift systematically with the microwave frequency, demonstrating that the additional structure is associated with the dispersive magnetic excitations. Below $T^{*}$, we fit each of the two features separately. Each fit uses a single line shape of the form of Eq. \ref{eq:2} plus a linear background, within a field window around that feature. Where the two features overlap, the tail of one mode extends into the fitting window of the other. The fit partly absorbs this tail into the line shape and background. This biases $H_r$, and especially $\Delta H$. To estimate this systematic error, we shift the window boundaries by about $\pm 200$ Oe and repeated the fits. We add the resulting spread to the statistical fit error in Figs.\ref{Figure 4}(b) and \ref{Figure 4}(c). The 14-GHz spectrum at 50 K is shown in Fig. \ref{Figure 4}(a), together with the corresponding fitted components. Two distinct resonance features are required to reproduce the experimental response, confirming the presence of a high-field resonance mode and a low-field resonance mode. The appearance of the additional mode below approximately 200 K coincides with the temperature at which substantial changes are observed in the extracted dynamic parameters. Therefore, the low-field mode is unlikely to originate simply from the measurement noise or fitting uncertainty, and instead indicates a modification of the magnetic state or magnetic energy at low temperature.
The temperature evolution of the FMR response is further quantified through the resonance field $H_{r}$ and the linewidth $\Delta H$, as summarized in Figs. \ref{Figure 4}(b) and (c). For clarity, the analysis is divided into two temperature regimes, separated by a characteristic temperature $T^{*} \approx 200$ K. Above this temperature, the spectra are dominated by a single resonance, whereas two distinct resonance components are resolved at lower temperatures. The temperature-dependent static magnetization M(T), is also included in the figures (magenta dashed line) for comparison. Notably, the emergence of the additional resonance is not accompanied by a comparably pronounced anomaly in the static magnetization, suggesting that the change in the dynamical response does not correspond to an obvious bulk magnetic transition. As shown in Fig. \ref{Figure 4}(b), the resonance field of the single mode decreases substantially upon cooling, from approximately 3200 Oe at 350 K to 2000 Oe at 200 K. This behavior is consistent with the increasing effective internal magnetic field associated with the development of ferromagnetic order and the evolution of magnetic anisotropy below ($T_C$ $\approx$ 335 K). Below ($T^{*}$), the two resonance branches exhibit distinctly different temperature dependence. The low-field mode continues to shift toward lower magnetic fields upon cooling, broadly following the trend of the high-temperature resonance. In contrast, the high-field mode remains within a comparatively narrow field range and exhibits only a weak, non-monotonic temperature dependence. This contrasting behavior suggests that the two resonances respond differently to the evolving magnetic environment. Although the low-field branch appears to follow the temperature evolution of the high-temperature resonance, the present measurements do not establish an unambiguous correspondence between these modes.

Similarly, the linewidth ($\Delta H$) is plotted and shown in Fig. \ref{Figure 4}(c). For a single mode, $\Delta H$ exhibits a  non-monotonic temperature dependence. Although it decreases overall on cooling, a pronounced broadening is also observed around 300-320 K. Since this temperature range lies relatively close to the ferromagnetic ordering temperature, the increased linewidth may be associated with increased magnetic fluctuations and relaxation processes near $T_C$. However, contributions from magnetic inhomogeneity and other extrinsic broadening mechanisms cannot be excluded, and the linewidth should therefore not be interpreted solely in terms of intrinsic Gilbert damping. Below $T^*$, the overlap affects the linewidth most. In an individual fit, the tail of the neighboring mode broadens the fitted line. This explains the scatter of ($\Delta H^\mathrm{high}-\Delta H^\mathrm{low}$) in Fig. S5(b). The low-field mode shows an apparent linewidth maximum near $\approx$ 150 K. This is where the two features are closest together. We therefore do not assign a physical meaning to this maximum. Below $T^*$, we use the linewidth only to show that both components are broad ($\Delta H \sim 0.5$-1.5 kOe). The resonance fields are more reliable. At 50-100 K, they shift smoothly with frequency [Fig. S5(a)].

\begin{figure*}
\centering
\includegraphics[width=1.0\textwidth]{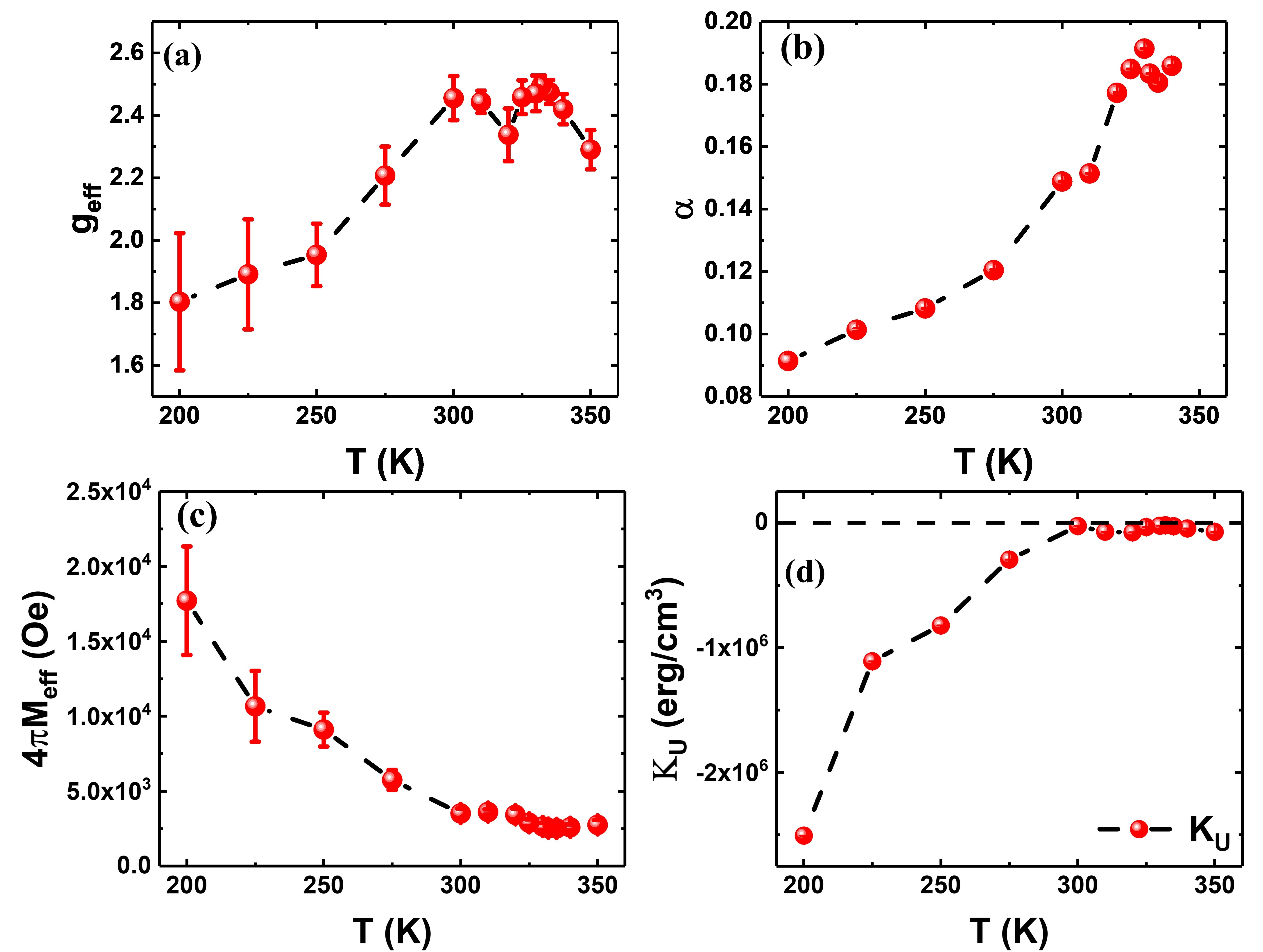}
\caption{Temperature dependence of the magnetic resonance parameters obtained from FMR measurements in the temperature range of 200–350 K: (a) effective Lande $g$-factor $(g_{{eff}})$, (b) damping constant $(\alpha)$, (c) effective magnetization $(4\pi M_{{eff}})$, (d) uniaxial magnetocrystalline anisotropy constant $(K_{U})$, evaluated from the FMR spectra as a function of temperature. Where not visible, the error bars are smaller than the symbol size. Dashed lines are guides to the eye. Points at 340 and 350 K ($T > T_C$) are effective parameters of a field-polarized state (see text).}
\label{Figure 5}
\end{figure*}

The temperature dependence parameters obtained from the FMR spectra are summarized in Fig. \ref{Figure 5}. The experiments are carried out over a temperature range of 350-25 K. However, below 200 K, the frequency range is too narrow for a reliable Kittel fit of either low- or high- field branches,  because $g_{eff}$ and $4\pi M_{eff}$ are strongly correlated (see below). We therefore evaluate $g_{eff}$, $\alpha$, $4\pi M_{eff}$ and $K_U$ only down to 200 K. Two of our measurement temperatures, 340 and 350 K, are above $T_C = 335$ K. At these temperatures, the applied field aligns the ferromagnetic correlations that remain above $T_C$. This field-polarized state precesses uniformly. Its resonance therefore still follows Eq. \ref{eq:3}. Above $T_C$, however, $M$ changes with the applied field. So $g_{eff}$ and $4\pi M_{eff}$ at 340 and 350 K are only effective values. Near $T_C$, critical spin fluctuations also broaden the line. So $\alpha$ in this range is an effective damping, not the true Gilbert damping. The effective $g$ factor ($g_{eff}$) is about 2.3 at 350 K. It reaches a maximum of about 2.45 $\pm$ 0.03 between 300 and 330 K. It then falls to about 1.8 $\pm$ 0.2 at 200 K [Fig. \ref{Figure 5}(a)] \cite{jiang2026contrasting}. This trend must be read with care. In the in-plane Kittel fit, $g_{eff}$ and $4\pi M_{eff}$ are strongly correlated \cite{shaw2013}. On cooling, the error in $g_{eff}$ grows strongly [Fig. \ref{Figure 5}(a)]. The overall decrease becomes particularly pronounced below approximately 300 K. Such a temperature dependence suggests that the effective gyromagnetic response evolves with the magnetic environment. In a material containing multiple inequivalent magnetic sites, changes in the relative contributions of different magnetic environments may influence the $g_{eff}$ obtained from a single-mode Kittel analysis. Above 200 K, where the spectra can be treated reasonably well as a single resonance mode, the frequency dependence of $H_{r}$ is fitted using the in-plane Kittel relation. The large departure of the effective g-factor from the spin-only value (g = 2) indicates that a simple picture of an isolated spin moment is inadequate, which is expected for a metallic Fe/Co-Te system with appreciable spin–orbit coupling \cite{jiang2026contrasting}. 

Linewidth analysis provides a complementary view of the temperature evolution. The relatively large damping close to $T_C$ is consistent with the broad FMR spectra in this temperature range and may reflect enhanced magnetic fluctuations together with extrinsic broadening associated with the chemically and structurally complex Fe/Co sublattice. Gilbert damping $(\alpha)$ gradually decreases from $\sim (0.18-0.19)$ at 340-320 K to $(\sim 0.096)$ at 200 K, indicating a substantial reduction in the contribution of measured frequency-dependent line width after cooling. This behavior suggests that the processes governing magnetic relaxation evolve considerably with temperature [Fig. \ref{Figure 5}(b)]. The relatively large damping values indicate substantial magnetic relaxation in comparison with many conventional low-damping ferromagnets, although identifying the dominant relaxation mechanism requires further investigation. The damping observed in CFGT is relatively large in comparison to both conventional metallic ferromagnets and several other van der Waals magnetic systems \cite{wu2023cobalt} \cite{shen2021multi}. At room temperature, we obtain an effective Gilbert damping factor $\alpha$ $\sim 0.15$, which decreases to approximately 0.096 at 200 K. For comparison, conventional in-plane metallic ferromagnets typically show damping values one to two orders of magnitude smaller: ($\alpha$ $\sim 0.007-0.013$) is commonly reported for permalloy (NiFe) \cite{pal2022short,zhao2016experimental}, while CoFeB films typically exhibit values of order (0.004-0.015) \cite{liu2011ferromagnetic,conca2013low}, depending on thickness, interfaces, and magnetic anisotropy. The damping in vdW ferromagnets spans a broader range. Insulating $Cr_{2}Ge_{2}Te_{6}$, for example, has been reported to exhibit ($\alpha \sim 10^{-3}-10^{-2}$) \cite{zeisner2019magnetic}\cite{xu2023electrical}, whereas bulk metallic $Fe_{5}GeTe_{2}$ shows a considerably larger of approx. 0.035 at 300 K, decreasing to about 0.007 at low temperature \cite{alahmed2021magnetism}. The substantially large $\alpha$ value obtained for Co-doped sample therefore appears to be characteristic of the strongly broadened magnetic response of the Co-doped crystal rather than a generic property of vdW ferromagnets. The metallic nature of the crystal, the appreciable spin-orbit coupling, and the presence of several inequivalent Fe/Co relaxation channels in the Fe environment may contribute to the measured damping \cite{lawati2025spin}. We therefore regard the large damping primarily as evidence for strong magnetic relaxation and a complex dynamical response in CFGT, rather than entirely assigning it to an intrinsic Gilbert relaxation mechanism. 

A particularly strong temperature dependence is observed in the effective magnetization, $4\pi M_{eff}$ which increases with cooling, from 2.7 $\pm$ 0.3 Oe at 350 K to 17 $\pm$ 3 Oe at 200 K [Fig. \ref{Figure 5}(c)], indicating a strong enhancement of the effective magnetization at lower temperatures. This pronounced enhancement indicates that the balance between magnetization and magnetic anisotropy contributions to the resonance condition changes significantly upon cooling.  Consequently, its temperature dependence reflects changes in the magnetic energy landscape rather than simply the growth of the magnetic moment. The magnetocrystalline anisotropy constant is expressed as: $K_{U}$ = $(M_{s}/2)(4\pi M_{s}-4\pi M_{eff})$, and its evolution with temperature is shown in Fig. \ref{Figure 5}(d). The calculated $K_{U}$ shows a negative sign at room temperature, and increases with lowering temperature. The predominantly negative values of $K_{U}$ are consistent with easy-plane magnetic anisotropy under the adopted sign convention \cite{may2019ferromagnetism}. While the anisotropy remains relatively weak at higher temperatures, its magnitude increases considerably upon cooling, reaching approximately (-2.5$\times$ 10$^6$) $erg/cm^{3}$ at 200 K. This evolution indicates a strengthening of the easy-plane anisotropy contribution relative to the magnetostatic term and highlights the substantial temperature dependence of the magnetic energy landscape in CFGT. Such behavior may arise from the combined temperature dependence of the magnetization, magneto-crystalline anisotropy, and magnetostatic contributions. Interestingly, the pronounced evolution of the anisotropy occurs as the system approaches the characteristic temperature $T^{*} \approx 200$ K, below which the FMR spectra develop an additional resonance mode. Although the present measurements do not establish a direct microscopic connection between these observations, the strong temperature dependence of the anisotropy suggests that changes in the magnetic energy landscape may contribute to the reconstruction of the low-temperature magnetization dynamics \cite{ni2026mesoscale}.

\subsection{\label{sec:level4}Conclusion}
In summary, we have investigated the temperature-dependent magnetization dynamics of CFGT using broadband ferromagnetic resonance spectroscopy. At high temperatures, the FMR response is dominated by a single resonance mode that can be described within the conventional Kittel framework, allowing the effective magnetization, spectroscopic $g$-factor, linewidth, and effective Gilbert damping to be determined. Upon cooling, these parameters evolve substantially, reflecting a pronounced temperature dependence of the magnetic energy landscape and relaxation processes in the material. 
A particularly striking change occurs below a characteristic temperature $T^{*} \approx 200$ K, where an additional resonance develops and the FMR response evolves into a two-mode regime. The additional feature persists to the lowest measured temperature and exhibits a systematic frequency dependence, indicating a genuine dynamical excitation rather than an artifact of spectral broadening. Notably, this dynamical crossover is not accompanied by a comparably pronounced anomaly in the static magnetization, highlighting the sensitivity of FMR to changes in the magnetic environment that are not apparent from bulk magnetometry. Although the microscopic origin of the additional mode cannot be uniquely determined from the present measurements, the complex and inequivalent Fe/Co magnetic environments of CFGT provide a plausible setting for the emergence of multiple dynamical responses. The magnetic anisotropy also exhibits a strong temperature dependence. The negative anisotropy constant is consistent with the easy-plane character of the crystal, while its increasing magnitude upon cooling indicates a substantial strengthening of the effective easy-plane anisotropy. The concurrent evolution of the anisotropy and the emergence of the multimode FMR response suggest that changes in the magnetic energy landscape may play an important role in the low-temperature dynamics. Overall, these results demonstrate that CFGT hosts a considerably richer temperature-dependent dynamical response than is evident from its static magnetic properties and establish broadband FMR as a sensitive probe of the magnetic interactions and relaxation processes in this high-$T_{C}$ van der Waals ferromagnet.

\subsection{Acknowledgement}%
This work was supported by the Ministry of Education, Singapore. S.A. acknowledges Jiang Luwen (National University of Singapore) for fruitful discussions. S.S. acknowledges Bindu (Hebrew University of Jerusalem, Israel) and Bivas Rana (Adam Mickiewicz University, Poznań, Poland) for valuable discussions and suggestions. S.S. also acknowledges financial support from the QUAKE project.

\clearpage  
\onecolumngrid  

\noindent \textbf{\LARGE Supplementary Information for "Temperature-Driven Evolution of Magnetization Dynamics in Cobalt-Doped $Fe_{5}GeTe_{2}$"}\\

\setcounter{figure}{0}
\renewcommand{\thefigure}{S\arabic{figure}}

\noindent \textbf{S1. Static Magnetization}:\\

\begin{figure*}[h!]
\centering
\includegraphics[width=0.8\textwidth]{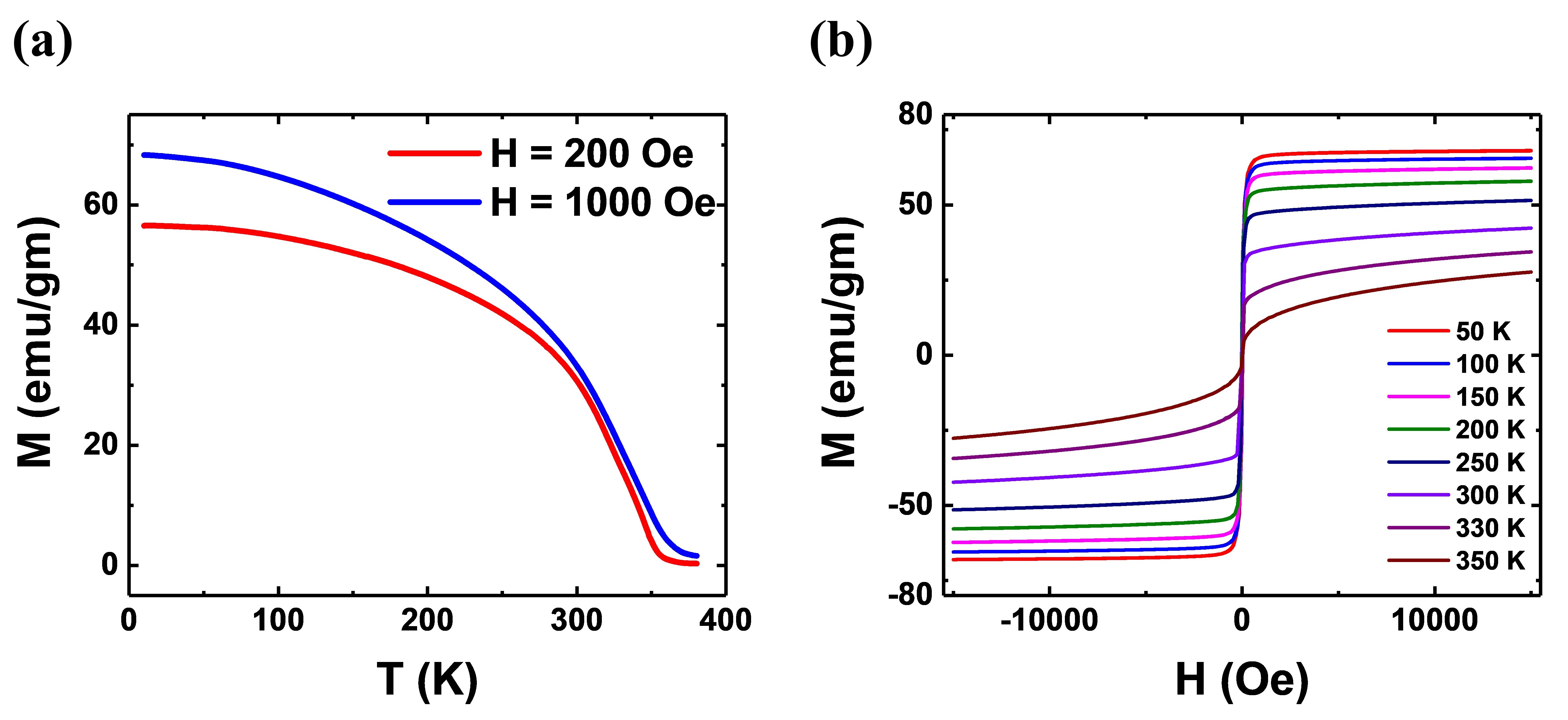}
\caption{(a) Temperature dependence of the magnetization M(T), measured under applied fields of H = 200 Oe (red) and 1000 Oe (blue). These two fields lie below and above the saturation field. The magnetization decreases steeply above about 300 K and becomes small near 350 K. The Curie temperature, $T_{C} \approx 335$ K, is taken from the minimum of $dM/dT$ [Fig. 1(b)]. (b)  Field sweep of magnetization M(H), measured over $\pm$ 15 kOe at temperatures from 50 K to 350 K. As the temperature rises, the saturation magnetization falls and the high-field slope grows, which shows the weakening ferromagnetic order as $T_C$ is approached.}
\label{SM_Figure1}
\end{figure*}

\clearpage

\noindent \textbf{S2. Magnetocrystalline Anisotropy Constant}:

\begin{figure*}[h!]
\centering
\includegraphics[width=1.0\textwidth]{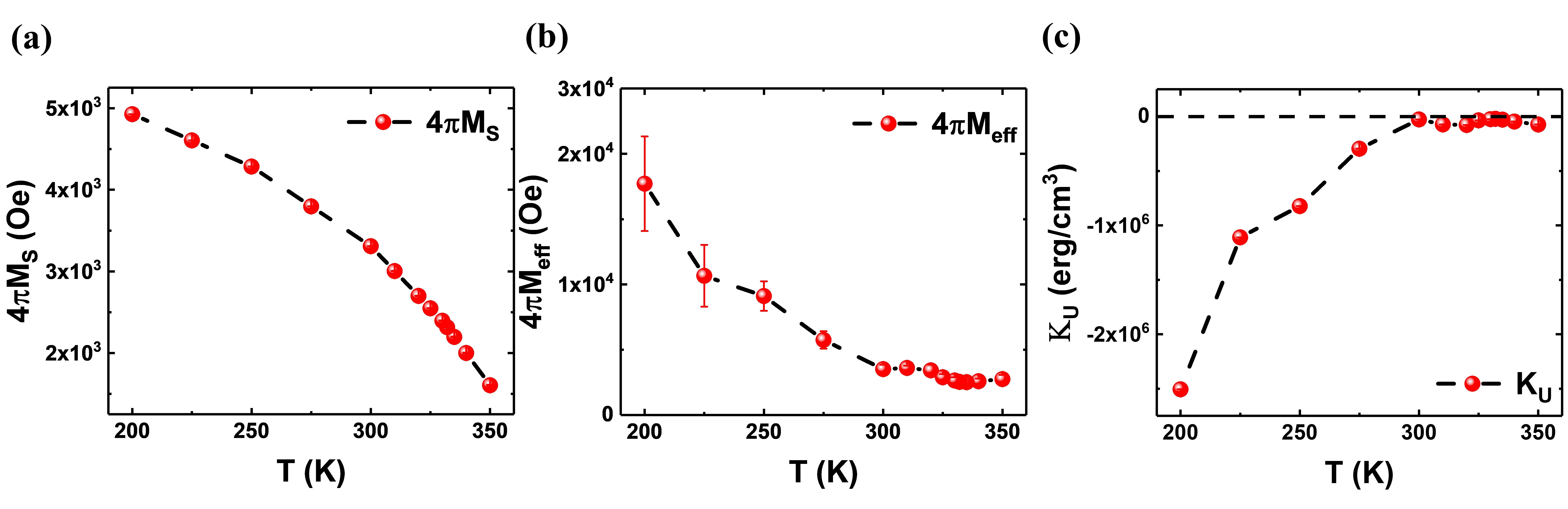}
\caption{Temperature dependence of the magnetic parameters of the sample, extracted over the range 200-350 K. (a) Saturation magnetization, $4\pi M_S$, which falls steadily from about 4.9 $\pm$ 0.0006 kOe at 200 K to 1.6 $\pm$ 0.006 kOe at 350 K. (b) Effective magnetization, $4\pi M_{eff}$, which drops from about 17 $\pm$ 3.6 kOe at 200 K and decreases at about 3 $\pm$ 0.2 kOe above 300 K. (c) Uniaxial magnetocrystalline anisotropy constant, $K_U$, obtained from $K_{U}$ = $(M_{s}/2)(4\pi M_{s}-4\pi M_{eff})$, is large and negative at low temperature, about $-2.5\times10^6$ erg/cm$^3$ at 200 K, and its magnitude shrinks quickly with rising temperature, reaching nearly zero (dashed line) above 300 K. Dashed lines through the data are guides to the eye.}
\label{SM_Figure2}
\end{figure*}

\clearpage

\noindent \textbf{S3. Temperature dependence of FMR spectra at 14 GHz}:

\begin{figure*}[h!]
\centering
\includegraphics[width=0.8\textwidth]{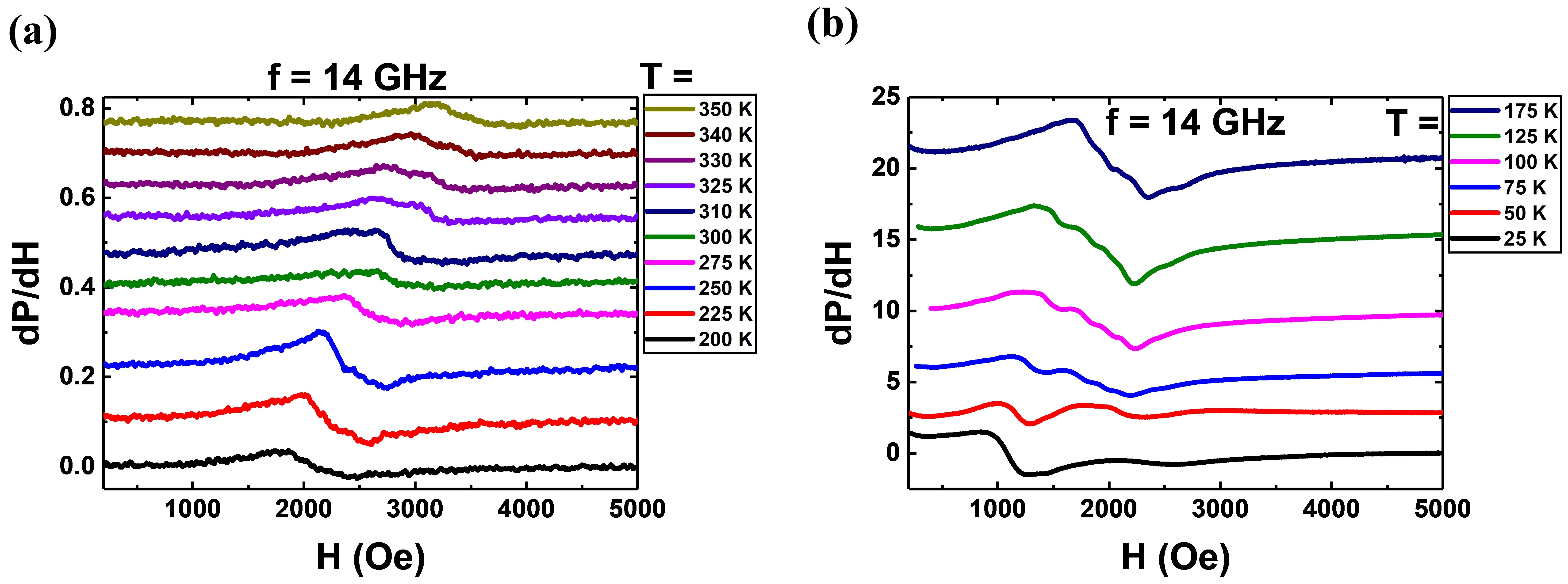}
\caption{Temperature evolution of the FMR response at 14 GHz, with spectra stacked and offset vertically for clarity. The same data appear as a temperature–field map in Fig. 3(a) of the main text. For clarity, the temperature range is split into two panels. (a) Field-derivative FMR signal measured at 14 GHz between temperature range 350-200 K. (b) Representative field-swept FMR spectra measured at 14 GHz between 175-25 K. Below 200 K, the additional mode is observed whose resonance field shifts towards lower fields as temperature reduces.}
\label{SM_Figure3}
\end{figure*}

\clearpage
\noindent \textbf{S4. Inhomogeneous Linewidth Broadening}:
\begin{figure}[h!]
\centering
\includegraphics[width=0.5\textwidth]{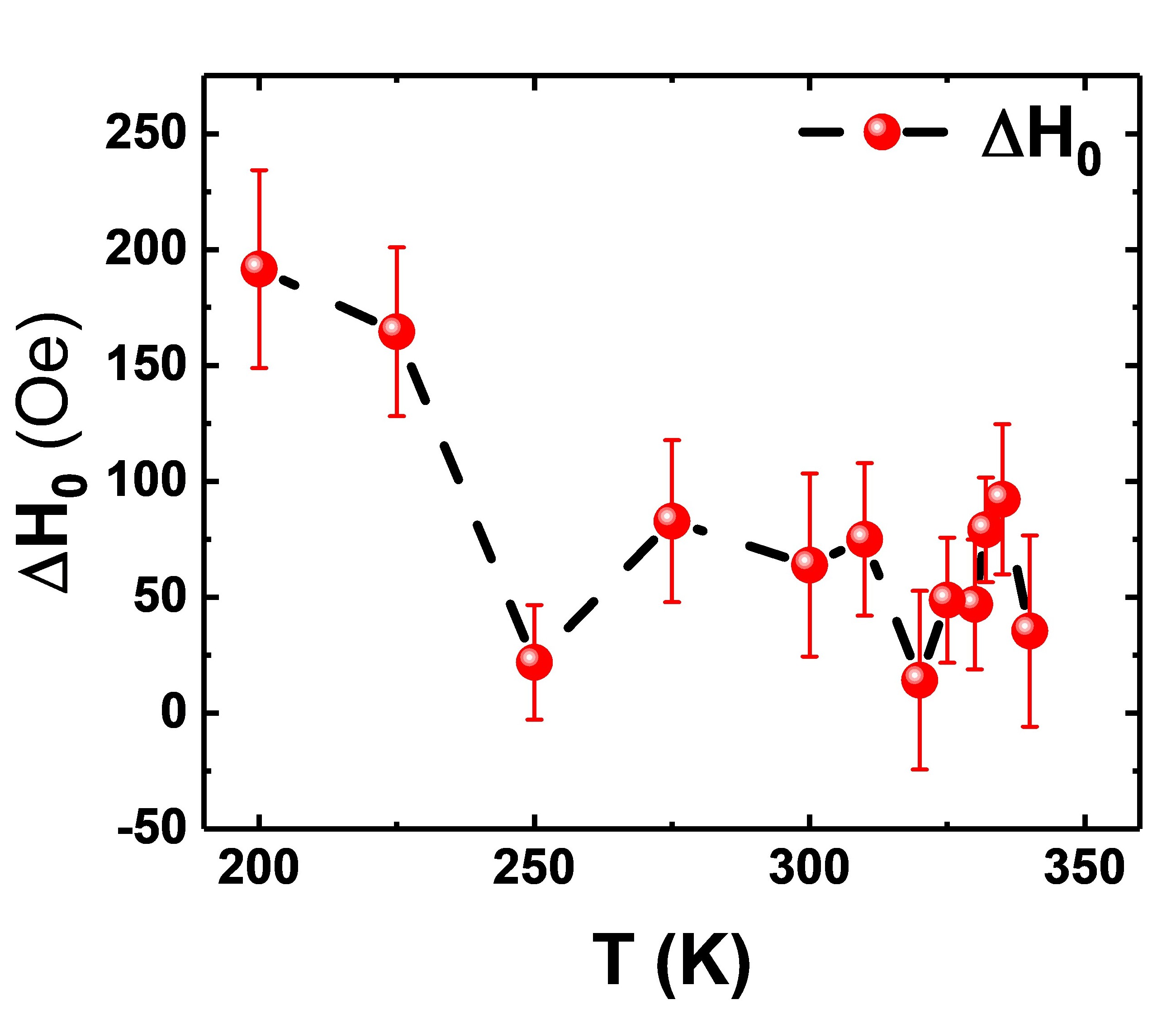}
\caption{Temperature dependence of the inhomogeneous linewidth broadening $\Delta H_0$ arising from a spatial distribution of the local magnetic anisotropy and magnetization. $\Delta H_0$ is calculated from the linear fit of linewidth versus frequency, $\Delta H= \frac{2h\alpha} {g\mu_{B}}f + \Delta H_{0}$ falls from about 190 Oe at 200 K to about 20 Oe at 250 K. Above 250 K it stays between roughly 15 and 95 Oe and shows no clear trend within the error bars. Error bars are from the linear fits, and the dashed line is a guide to the eye.}
\label{SM_Figure4}
\end{figure}

\clearpage
\noindent \textbf{S5. Separation of the two resonance modes below $T^{*}$}:

\begin{figure*}[h!]
\centering
\includegraphics[width=1.0\textwidth]{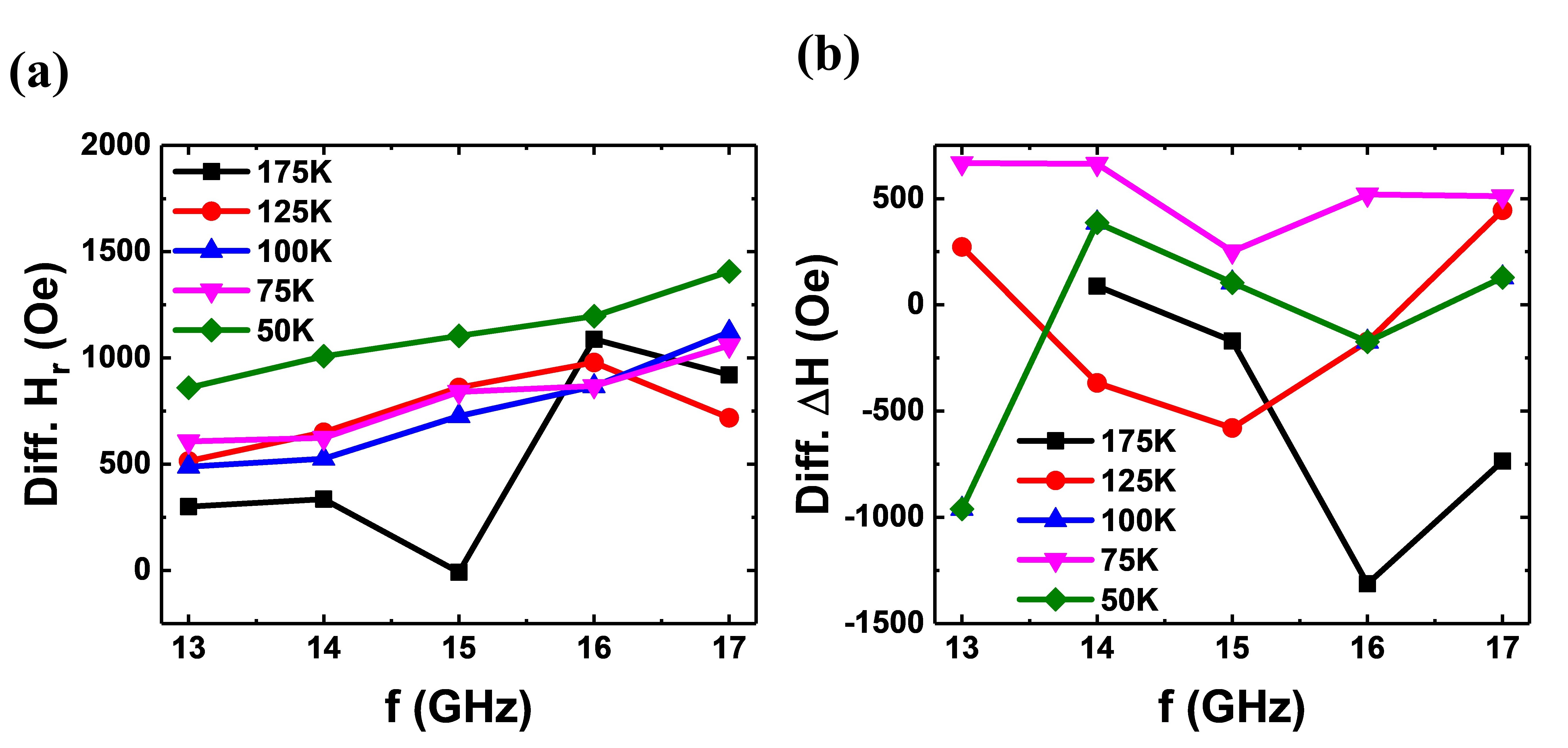}
\caption{Frequency dependence of the difference between the high-field and low-field FMR modes at temperatures from 50 to 175 K. (a) Difference in resonance field, ${Diff.}, H_r = H_r^\mathrm{high} - H_r^\mathrm{low}$. The mode separation increases with frequency and is largest at 50 K, rising from about 750 Oe at 12 GHz to about 1800 Oe at 17.5 GHz. At 175 K, where the two modes are weakest and partly overlap, the trend is less clear. (b) Difference in linewidth, $\Delta H^\mathrm{high}-\Delta H^\mathrm{low}$, which scatters between about $-1300$ and $+650$~Oe without systematic dependence on frequency or temperature. This reflects the bias from the overlapping tail of the neighbouring mode in the individual fits, so the individual linewidth are not well constrained when the modes overlap.}
\label{SM_Figure5}
\end{figure*}
\clearpage

\bibliography{references}

\end{document}